\documentclass[12pt,epsfig]{article}

\usepackage{amssymb,epsfig}

\newcommand{\be}{\begin{equation}}

\newcommand{\ee}{\end{equation}}

\newcommand{\bea}{\begin{eqnarray}}

\newcommand{\eea}{\end{eqnarray}}

\begin{document}

\begin{titlepage}

\begin{flushright}
\begin{tabular}{l}
 hep-ph/0211nnn
\end{tabular}
\end{flushright}
\vspace{1.5cm}

\begin{center}

{\LARGE \bf  Deeply Virtual Compton Scattering on Spin-1 Nuclei\footnote{to be
published in the Proceedings of the XVIth International Conference on
Particles and Nuclei
(PANIC02), Osaka, Japan, 30 September  4 October 2002}}
\vspace{1cm}

{\sc F.~Cano}${}^1$ and
{\sc B.~Pire}${}^{2}$,

\vspace*{0.1cm} ${}^1${\it
   DAPNIA/SPhN, CEA--Saclay, F91191
Gif-sur-Yvette Cedex, France
                       } \\[0.2cm]
\vspace*{0.1cm} ${}^2$ {\it
CPhT, {\'E}cole Polytechnique, F-91128 Palaiseau, France\footnote{
  Unit{\'e} mixte C7644 du CNRS.}
                       } \\[0.2cm]

\vskip2cm
{\bf Abstract:\\[10pt]} \parbox[t]{\textwidth}{
 We consider the Generalized Parton Distributions  for
spin-1 nuclei in general and on the deuteron in particular.
We use the impulse approximation to obtain a convolution
model for them. Sum rules are used to check the validity of
the approach and to estimate the importance of higher
Fock-space states in the deuteron. Numerical predictions for
the Beam Spin Asymmetry in deeply virtual Compton scattering
are presented. }
\vskip1cm
\end{center}

\vspace*{1cm}

\end{titlepage}

\section{Introduction}
Hard exclusive processes, such as deeply virtual Compton scattering (DVCS) and
deeply exclusive meson production (DEMP), have been recently demonstrated to
open the possibility of obtaining a quite complete picture of the hadronic
structure. The information which can be accessed through these experiments is
encoded by the Generalized Parton Distributions, GPDs\cite{GPD} (for recent
reviews see\cite{Guichon}). The physical interpretation of
the GPDs has been elucidated by some authors
\cite{BURKARDT00}. Recent measurements of the azimuthal
dependence of the beam spin asymmetry in DVCS
\cite{HermesCEBAF} have provided
experimental evidence to support the validity of the formalism of GPDs and the
underlying QCD factorization theorems.

	The formalism of GPDs can be applied to the deuteron as well
\cite{BERGER01}.
 From the theoretical viewpoint, it is the simplest
and best known nuclear system and represents the most appropriate starting
point to investigate hard exclusive processes off nuclei\cite{KIRCHNER02}.
Moreover, hard exclusive processes could offer a new source of information
 about the partonic degrees of freedom in nuclei, complementary to the existing
one from deep inelastic scattering.

A parameterization of the non-perturbative matrix
elements which determine the amplitudes in DVCS and DEMP on a spin-one
target were given in terms of nine GPDs for the quark
sector\cite{BERGER01} (five coming from the vector operator and four from the
 axial vector one).  Due to the spin-one
character of the target, there are more GPD's than in the nucleon case,
but at the same time the set of polarization observables which in
principle could be measured is also richer.

\begin{figure}
\includegraphics[scale=0.6]{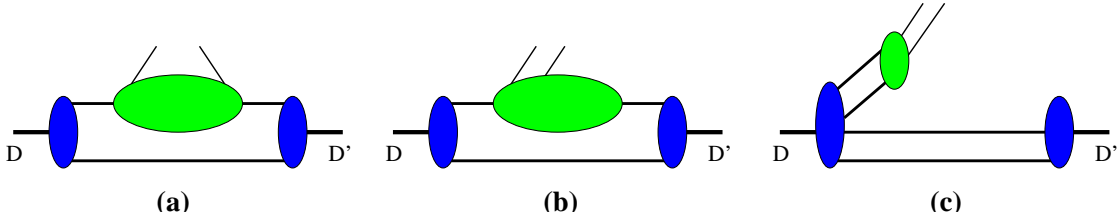}
\caption{Contributions to the deuteron GPDs in the impulse
approximation for the lowest $|pn\rangle$ Fock-space state of the
deuteron. The figure (c) corresponds to higher components that we have
neglected.}
\end{figure}

\section{Deuteron GPD's in the impulse approximation}

The simplest way to model deuteron GPDs is to use the impulse approximation
where the interaction with photons occurs in a single nucleon the other
being a spectator (see fig. 1). For the sake of simplicity we will focus in
the following  on the helicity independent GPD's but analogous relations
can be found for the helicity dependent ones. Since the deuteron is an
isoscalar target we have:

\begin{equation}
H_i^u(x,\xi,t) = H_i^d(x,\xi,t) \equiv H_i^q(x,\xi,t)
\end{equation}

	The relationship between deuteron GPDs and the helicity matrix
elements of the non-local quark vector operator is given by:

\begin{equation}
H_i^q(x,\xi,t) = C_i^{\lambda '\lambda} V^q_{\lambda ' \lambda}
\end{equation}

\noindent where $C_i^{\lambda ', \lambda}$ are
coefficients which depends on the polarization vectors of the deuteron and
on the choosen kinematics and $V^q_{\lambda
' \lambda}$ is given  in the impulse approximation  by a convolution
between deuteron wave functions and the isoscalar combination of nucleon
GPDs:

\begin{eqnarray}
V^q_{\lambda ' \lambda} & =  & \frac{2}{(16\pi^3)}\int \; d\alpha \,
d\vec{k}_\perp \,
 \sqrt{\frac{1+\xi}{1-\xi}} \frac{1}{\sqrt{\alpha \alpha '}}
\sum_{\lambda_1 ',\lambda_1,\lambda_2} \chi^*_{\lambda
'}(\alpha',\vec{k}_\perp ',\lambda_1 ',\lambda_2)
\chi_{\lambda}(\alpha,\vec{k}_\perp,\lambda_1,\lambda_2)
\nonumber \\ & \cdot & \left[ 2 (\sqrt{1-\xi_N^2} H^{\mbox {\scriptsize
IS}}(x_N,\xi_N,t) -
\frac{\xi_N^2}{\sqrt{1-\xi_N^2}} E^{\mbox {\scriptsize IS}}(x_N,\xi_N,t))
\delta_{\lambda_1 ' \lambda_1,} \right. \nonumber \\
&  & \;\;\;\;\;\;+\left. 2 \frac{\sqrt{t_0 - t}}{2 M_N} \eta_{\lambda_1}
E^{\mbox {\scriptsize IS}}(x_N,\xi_N,t)
\delta_{\lambda_1 ',-\lambda_1,} \right]
\end{eqnarray}

\begin{figure}
\includegraphics[scale=.6]{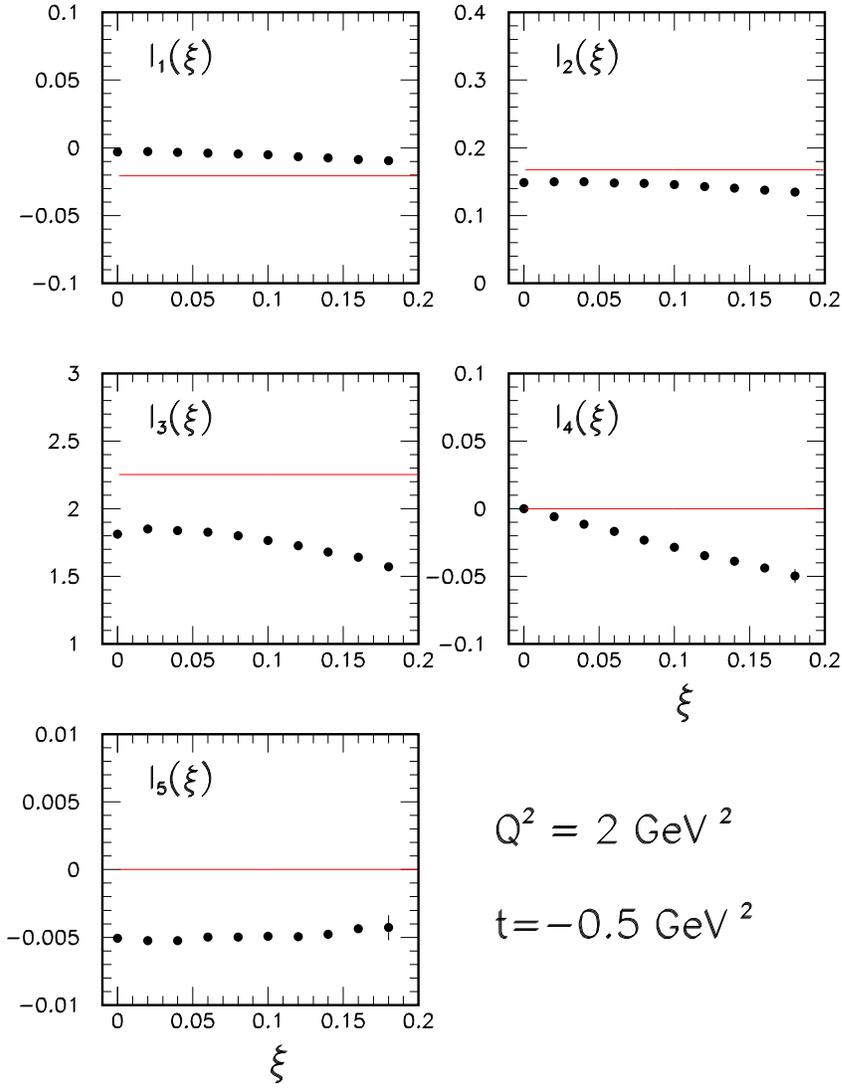}
\caption{$\xi$ dependence of the first moments of vector GPDs
$H_i$. Lines correspond to the theoretical
($\xi$-independent) expected value according to the sum rules and
points are the actual values obtained with the impulse
approximation for the GPDs.}
\end{figure}

\noindent
where $\alpha$ refers to the fraction of plus momentum carried by the
active nucleon in the initial deuteron and
$\vec{k}_\perp$ to its transverse momentum in a frame where
$\vec{P}_\perp=0$.  The kinematics of the process imposes
that $\alpha ' = \frac{\alpha (1+\xi) - 2 \xi }{1-\xi}$ and $ \vec{k}_\perp
' = \vec{k}_\perp - \left(
\frac{1-\alpha}{1-\xi } \right) \vec{\Delta}_\perp$. The integral over
$\alpha$ is appropiately bounded from below to
ensure the positivity of plus momentum involved in the problem.

The combinations of nucleon GPDs which appear are the isoscalar ones:
$
H^{\mbox {\scriptsize IS}}(x_N,\xi_N,t)=\frac{1}{2}(H^{u}(x_N,\xi_N,t)+
H^{d}(x_N,\xi_N,t))
$
with $\xi_N  =  \frac{\xi}{\alpha (1+ \xi)-\xi}  \;\; ,
x_N  =  \frac{x}{\xi} \;  \xi_N $ .
The minimal
value of the momentum transfer is $t_0 = - \frac{4 M_D^2
\xi^2+\vec{\Delta}_\perp^{\, 2}}{1 -
\xi^2}$ and
$\eta_{\lambda_1}$ is a phase.

It is clear that for most cases the dominant contribution will be the one
proportional to $H^{\mbox {\scriptsize IS}}$.  Numerically, the term that
goes with  $E^{\mbox {\scriptsize IS}}$  has
little effect on the cross sections.

In the impulse approximation we have discarded higher Fock-space states in
the deuteron (see fig. 1). An important check of our model is the
$\xi$-independence of the integrated
quantities $\int_{-1}^1 H_i(x,\xi,t)$ at fixed $t$. These sum rules relate
the x-integrated GPD's to
the form factors of local vector and axial currents.
They read:
\begin{equation}
 \int_{-1}^1 H_i(x,\xi,t) = G_i (t) \;\;\;\;\;\;(i=1,2,3)
\;\;\;\;\;\;\;\;\;\; ; \;\;\;\;\;\;\;\;\int_{-1}^1
H_i(x,\xi,t) = 0
\;\;\;\;\;\;(i=4,5)
\end{equation}
We see on fig. 2, where we have plotted $I(\xi) \equiv \int_{-1}^1
H_i(x,\xi,t=-0.5$GeV$^2)$,
that these sum rules are quite well verified by our model. The variation of
$I_i(\xi)$ with
$\xi$ gives a measure of the physical ingredients which are missing in the
impulse
approximation.

\section{Beam spin asymmetry in DVCS}
Let us now present one observable calculated with our modelized deuteron
GPD's, namely the beam spin asymmetry in DVCS. It is defined as

\begin{equation}
A_{LU} (\phi) = \frac{d\sigma^\uparrow (\phi) - d\sigma^\downarrow
(\phi)}{d\sigma^\uparrow (\phi)+ d\sigma^\downarrow
(\phi)}
\end{equation}
where $\phi$ is the angle between the lepton and hadron scattering planes.
The numerator is proportional to the
interference between the Bethe-Heitler and the DVCS amplitudes. A very
rough approximation, in the case of the
dominance of the Bethe-Heitler process, leads to an asymmetry proportional
to $\sin(\phi)$.

Our predictions are shown on Fig. 3 for JLab and Hermes energies. The sign
of the asymmetry is reversed for a
positron beam. Such a sizable asymetry should be quite easily measured. It
will constitute a crucial test of the validity of our model.

\begin{figure}
\includegraphics[scale=.6]{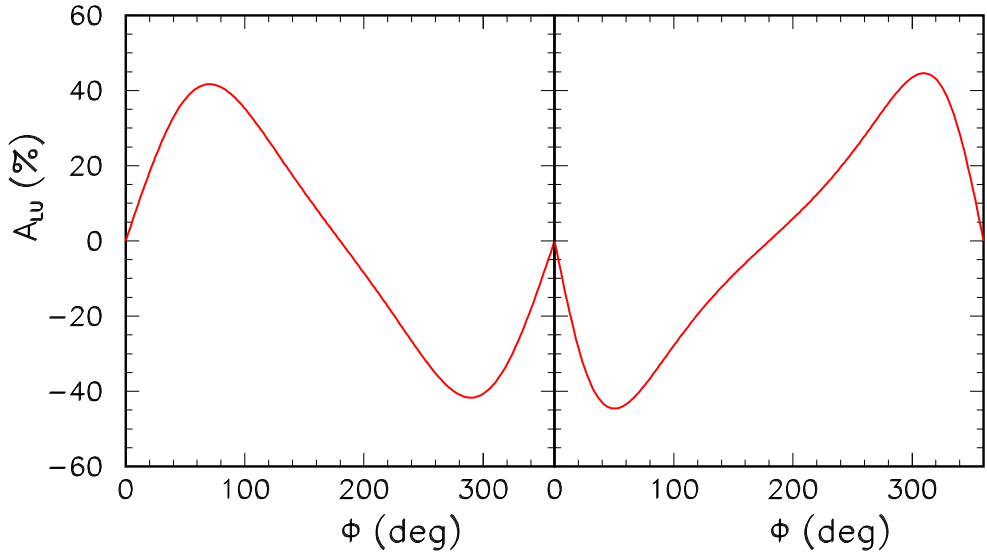}
\caption{Azimuthal dependence of the Beam Spin Asymmetry as defined in
the text. Left: $x_{\mbox {\tiny Bj}}=0.2$, $Q^2= 2$ GeV$^2$ and $E_e=
6$ GeV. Right: $x_{\mbox {\tiny Bj}}=0.1$, $Q^2= 4$ GeV$^2$ and $E_{e^+}=
27$ GeV. In both cases $t$ is fixed to $-0.3$ GeV$^2$.}
\end{figure}

\vspace*{1cm}

{\bf Acknowledgements}

\noindent
This work has been supported by the EC--IHP Network ESOP,
Contract HPRN-CT-2000-00130.

\end{document}